\documentclass[aps,prl,twocolumn,groupedaddress,showpacs]{revtex4}

\usepackage{graphicx}
\usepackage{amsfonts}
\usepackage{amsmath}
\usepackage{amssymb}
\usepackage{bm}
\begin{document}
%\begin{CJK}{GBK}{song}
%\title{Making quantum spin-Hall effect robust against time-reverse-symmetry-broken perturbations}
\title{Stabilization of Quantum Spin Hall Effect by Designed Removal of
    Time-Reversal Symmetry of Edge States}
\author{Huichao Li$^1$}
\author{L. Sheng$^1$}
\email{shengli@nju.edu.cn}
\author{R. Shen$^1$}
\author{L. B. Shao$^1$}
\author{Baigeng Wang$^1$}
\author{D. N. Sheng$^2$}
\author{D. Y. Xing$^1$}
\email{dyxing@nju.edu.cn}
\affiliation{$^1$National Laboratory of Solid State Microstructures and
Department of Physics, Nanjing University, Nanjing 210093, China\\
$^2$ Department of Physics and Astronomy, California State
University, Northridge, California 91330, USA}
\date{\today }

\begin{abstract}
The quantum spin Hall (QSH) effect is known to be unstable to perturbations
violating time-reversal symmetry.
%While it is known to be fragile in the presence of perturbations
%violating time-reversal symmetry, we show that the quantum spin Hall
%(QSH) effect can be stablized by proper magnetic manipulation.
We show that creating a narrow ferromagnetic (FM) region near
the edge of a QSH sample can push one of the counterpropagating edge
states to the inner boundary of the FM region, and leave the other
at the outer boundary, without changing their spin polarizations and
propagation directions. Since the two edge states
are spatially separated into different ``lanes'', the QSH
effect becomes robust against symmetry-breaking perturbations.
\end{abstract}

 \maketitle

The quantum spin Hall (QSH) effect, a quantum state of matter,
has attracted much attention in recent years, because of its
fundamental interest and potential applications in spintronic
devices. The QSH effect was first predicted theoretically by Kane
and Mele~\cite{qshe1} and by Bernevig and Zhang~\cite{qshe2}, in
independent works. Soon after,
the QSH effect was observed experimentally in HgTe
quantum wells,~\cite{HgTe} following theoretical prediction.~\cite{HgTe0}
The discovery of the QSH effect has inspired the
theoretical proposals~\cite{3dti1,3dti2,3dti3,3dti4,3dti5}
for topological insulators in three dimension, which have been
confirmed experimentally.~\cite{Exp3D1,Exp3D2,Exp3D3,Exp3D4,Exp3D5,Exp3D6}
A key ingredient to the QSH effect
is a strong intrinsic spin-orbit coupling, which acts as
spin-dependent magnetic fluxes coupled to the electron momentum. In
the ideal case, where electron spin is conserved, the two spin
sectors of a QSH system behave like two independent quantum Hall
(QH) systems without Landau levels.~\cite{haldane} They contribute
opposite quantized Hall conductivities, when the electron Fermi
level is inside the bulk band gap, so that the total Hall
conductivity vanishes but the spin Hall conductivity is quantized.
On a sample edge, two counterpropagating gapless edge modes with
opposite spin polarizations exist in the bulk band gap, which can
transport spin currents without dissipation of energy.

When the spin conservation is destroyed, e.g., by the Rashba-like
spin-orbit coupling, the spin Hall conductivity deviates from the
quantized value.~\cite{LSEdge} However, the edge transport can
remain to be dissipationless,~\cite{qshe1,helical} provided that the
time-reversal (TR) symmetry is present and the bulk band gap is not
closed. In this case, a QSH system can no longer be divided into two
QH systems, and the existence of the gapless edge states has been
attributed to the nontrivial topological properties of bulk energy
bands. The nontrivial bulk band topology of the QSH systems is
usually described by the $Z_2$ index~\cite{Z2} or the spin Chern
numbers.~\cite{spinch1,spinch2,spinch3} 
%These topological invariants
%reveal the fundamental distinction between a QSH insulator and an
%ordinary band insulator.

%Extension of the idea of the QSH effect to higher dimension has led
%to the proposals of three-dimensional (3D)
%TIs.~\cite{3dti1,3dti2,3DTI_Model,3d1,3d2} Research of the 3D TIs
%has been fruitful in recent years, both theoretically and
%experimentally. A 3D TI has a bulk band gap and gapless surface
%states on the sample boundary. A simple 3D TI consists of layers of
%2D QSH systems. However, such a 3D TI is not stable in the presence
%of disorder due to interlayer scattering, and so called the weak TI.
%A strong TI is topologically nontrivial in any direction, and the
%surface states are characterized by an odd number of Dirac cones.
%The metallic surface states of the 3D TIs provide a unique platform
%for realizing some exotic physical phenomena, such as Majorana
%fermions~\cite{Majorana} and topological magnetoelectric
%effect.~\cite{Magneto1,Magneto2} The existence of topological
%surface states in the 3D TIs has been experimentally confirmed in
%Bi$_{1-x}$Sb$_x$, Bi$_2$Te$_3$, and Bi$_2$Se$_3$
%materials,~\cite{Exp3D1,Exp3D2,Exp3D3,Exp3D4,Exp3D5,Exp3D6} which
%evokes a great surge of research interest in this field.

While the TR symmetry was often considered to be a prerequisite to
the QSH effect, its role is two-sided. In a TR invariant QSH system,
the two oppositely moving edge states at the Fermi energy are
connected to each other under TR, and so have opposite spin
orientations. Elastic backscattering from nonmagnetic random
potential is forbidden. On the other hand, the two opposite movers
have identical spatial probability distributions. Turning on small
TR-symmetry-breaking perturbations immediately couples the two edge
states, giving rise to backscattering. This makes the QSH effect
fragile in realistic environments, where perturbations violating the TR
symmetry are usually unavoidable. Experimentally, two-terminal
conductance close to the predicted value $2(e^2/h)$
was observed only for small QSH samples with dimensions of
about $(1\times 1)\mu m^2$,~\cite{HgTe} in contrast to the
traditional QH effect, where the Hall conductivity can be
precisely quantized on macroscopic scales.
So far, QSH
effect as robust as the QH effect has been elusive.
It was found recently that the nontrivial bulk band
topology of the QSH systems remains intact, even when the TR
symmetry is broken,~\cite{spinch4} implying that the instability of
the QSH effect is solely due to properties of the edge states.

In this Letter, we show that the QSH effect can be
stabilized in two-dimensional topological insulators by 
inducing ferromagnetism on narrow strips
along the edges. As a result of the quantum anomalous Hall (QAH) effect 
generated by the exchange field~\cite{HgTe2} within a ferromagnetic (FM) strip,
one of the helical edge states is pushed to the inner
boundary of the FM region, and the other remains on
the outer boundary. The
moving directions and spin orientations of the individual edge
states are unchanged, so that the QSH effect persists. Importantly,
the edge states are spatially separated, so that the QSH effect
becomes robust against general perturbations without fictitious
symmetry constraints. We present both qualitative discussion
and quantitative calculation to demonstrate the physical picture and
practical feasibility of this proposal.
Our work paves a road to realize robust QSH
effect via magnetic manipulation.

We start from the effective Hamiltonian for a HgTe quantum
well~\cite{HgTe0} with an exchange field given by $H=H_{0}+H_{1}$
with
\begin{equation}
H_{0}=v_{\mbox{\tiny
F}}(\hat{\tau}_zk_x\hat{\sigma}_x+k_y\hat{\sigma}_y)+Dk^2
+(M_{0}-Bk^2)\hat{\sigma}_z\ .
\end{equation}
Here, $v_{\mbox{\tiny F}}$, $D$, $M_{0}$ and $B$ are the parameters
of the model, $\hat{\mbox{\boldmath{$\tau$}}}$ stand for Pauli matrices for two
spin states, and $\hat{\mbox{\boldmath{$\sigma$}}}$ for the electron and
hole bands. An exchange field can be created in the HgTe layer by
doping of magnetic atoms, such as Mn.~\cite{HgTe2} Within the mean-field
approximation, the exchange field
can be described by~\cite{HgTe2}
$H_{1}=(g_{0}\hat{\sigma}_z+g_{1})\hat{\tau}_z$, where
$g_{0}=\frac{1}{2}(G_{\mbox{\tiny H}}-G_{\mbox{\tiny E}})$ and
$g_{1}=\frac{1}{2}(G_{\mbox{\tiny H}}+G_{\mbox{\tiny E}})$ with
$2G_{\mbox{\tiny E}}$ ($2G_{\mbox{\tiny H}}$) as the exchange
splitting of the electron (hole) bands.
For convenience, we set the reduced Planck constant $\hbar$ to be unity.
$B^2>D^2$ is assumed to ensure the valence bands to be inverted.~\cite{spinch3,Shan}

%It is worth mentioning that
%the model Hamiltonian Eq.\ (1) can describe not only the HgTe
%quantum wells, but also a class of QSH
%systems.~\cite{HgTe3} For example, setting $g_{1}=0$ and
%making a unitary transformation ${\cal H}=U^\dagger HU$ with
%$U=\frac{1}{2}[(1+\hat{\tau}_z)
%+(1-\hat{\tau}_z)\hat{\sigma}_y]e^{-i\frac{\pi}{4}\hat{\sigma}_z}$, we
%obtain ${\cal H}=v_{\mbox{\tiny F}}(k_y\hat{\sigma}_x-k_x\hat{\sigma}_y)
%+\left(M_{0}-Bk^2\right)\hat{\tau}_z\hat{\sigma}_z+Dk^2
%+g_{0}\hat{\sigma}_z$. One can find that
%this Hamiltonian is identical to that of a thin film of 3D TI
%Bi$_{2}$Se$_{3}$ with an exchange field,~\cite{spinch3,Shan} for which
%$\hat{\mbox{\boldmath{$\sigma$}}}$ stand for spin up and down, and
%$\hat{\mbox{\boldmath{$\tau$}}}$ for bonding and antibonding of the
%surface states on two surfaces.

Since $\hat{\tau}_z$ is a conserved quantity, one can easily
diagonalize Eq.\ (1), and obtain two conduction bands and two
valence bands. Under the condition of $\vert
g_1\vert<\mbox{max}(\vert M_0\vert,\vert g_0\vert)$, a nonzero
middle band gap exists, except at $g_{0}=\pm M_{0}$ where the
conduction and valence bands touch; otherwise the conduction and
valence bands overlap and the system is a metal. As has been
discussed in Ref.\ \cite{HgTe2}, for HgTe quantum wells doped with
Mn, $G_{\mbox{\tiny E}}$ and $G_{\mbox{\tiny H}}$ have opposite
signs, so that the above condition is satisfied. Given $\vert
g_1\vert<\mbox{max}(\vert M_0\vert,\vert g_0\vert)$, the spin Chern
numbers for $\tau_{z}=\pm 1$ can be derived to be
$C_{\pm}=\pm\frac{1}{2}[\mbox{sgn}(B)+\mbox{sgn}(M_{0}\pm g_{0})]$.
At $g_{0}=0$, $C_{\pm}=\pm\mbox{sgn}(B)$ if $BM_{0}>0$,
corresponding to a QSH phase, and $C_{\pm}=0$ if $BM_{0}<0$,
corresponding to an ordinary insulator. We focus on systems with
$BM_{0}>0$, and for the sake of definiteness, we will
confine ourselves to the parameter region of  $B<0$ and $M_{0}<0$,
which is the case with the HgTe quantum wells exhibiting the QSH effect.
(All the conclusions reached in this work do also apply to $B>0$ and $M_{0}>0$.)
In this case, $C_{\pm}=(-1, 1)$ at $g_{0}=0$.
With increasing $g_0$ to $g_0=\vert M_0\vert$, $C_{\pm}$
undergo a transition from $(-1, 1)$ to $(0, 1)$, the latter
corresponding to a QAH phase.~\cite{spinch4,HgTe2}

\begin{figure}
\includegraphics[width=2.4in]{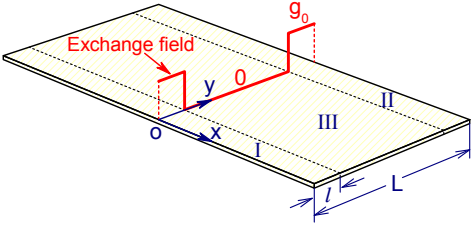}
\caption{Schematic of the QSH sample with a long strip geometry. The
profile of the $y$-dependent exchange field is shown by the thick
(red) line.} \label{Fig1}
  \end{figure}
Next we consider a QSH sample with a strip geometry, as shown in
Fig.\ 1. Since $C_{\pm}$ do not depend on $D$
and $g_{1}$ for $\vert g_1\vert<\mbox{max}(\vert M_0\vert,\vert
g_0\vert)$, without loss of generality, we set $D=g_{1}=0$ for now
to make a physical discussion, and the effect of finite $D$ and
$g_1$ will be taken into account in numerical calculation later. The
exchange field $g_{0}$ is taken to be nonzero in region I of
$0<y<l$ and region II of $(L-l)<y<L$, and vanishing in region III of
$l<y<(L-l)$. The system as a whole has a bulk energy gap around
energy $0$, for $g_{0}\neq\pm M_{0}$. The edge states in the bulk
energy gap can be solved analytically by replacing $k_y$ with
$-i\partial_y$ in the system Hamiltonian. For $\tau_z=1$, an edge
mode with energy $E_{+}(k_x)=v_{\mbox{\tiny F}}k_x$ is found on one
side of the strip, whose wavefunction is given by
\begin{equation}
\varphi_{+}(k_x,y)=\vert 1,1\rangle\phi_{+}(k_x,y)\ .
\label{phiP}
\end{equation}
Here, ket $\vert\tau_z,\sigma_x\rangle$ with $\tau_z=\pm 1$ and
$\sigma_x=\pm 1$ is used to represent the common eigenstate of
$\hat{\tau}_z$ and $\hat{\sigma}_x$. The spatial wavefunction
$\phi_{+}(k_x,y)=C[e^{-y/\xi_{1}(g_{0})}-e^{-y/\xi_{2}(g_{0})}]$ for $y<l$,
and $\phi_{+}(k_x,
y)=D_{1}e^{-(y-l)/\xi_{1}(0)}-D_{2}e^{-(y-l)/\xi_{2}(0)}$ for $y\ge
l$, with $C$, $D_{1}$ and $D_{2}$ to be determined from the
conditions of continuity and normalization of $\phi_{+}(k_x, y)$.
The two characteristic length functions are defined as
\begin{equation}
\xi_{1,2}(\epsilon)=\frac{2\vert B\vert}{
v_{\mbox{\tiny F}}\pm\sqrt{v_{\mbox{\tiny F}}^2-4B(M_0-Bk_x^2+\epsilon)}}\ .
\end{equation}
For $\tau_z=-1$, we find another edge mode with energy
$E_{-}(k_x)=-v{\mbox{\tiny F}}k_x$ and wavefunction
\begin{equation}
\varphi_{-}(k_x,y)=\vert -1, 1\rangle\phi_{-}(k_x,y)\ ,
\label{phiM}
\end{equation}
 where $\phi_{-}(k_x,y)=E[e^{-y/\xi_{1}(-g_{0})}-e^{-y/\xi_{2}(-g_{0})}]$ for
$y<l$, and $\phi_{-}(k_x,y)=F_{1}e^{-(y-l)/\xi_{1}(0)}-F_{2}
e^{-(y-l)/\xi_{2}(0)}$ for $y\ge l$. Owing to the two-fold rotation
symmetry, the edge modes on the other side of the strip have
dispersion relations $E_{\pm}(k_x)=\mp v_{\mbox{\tiny F}}k_x$.
%Their
%wavefunctions can be obtained through replacements $y\rightarrow
%L-y$, $\sigma_x\rightarrow -\sigma_x$ and $k_x\rightarrow -k_x$ in
%Eqs.\ (\ref{phiP}) and (\ref{phiM}).
%The energy spectrum of edge
%states appears to have mirror symmetry between $k_{x}$ and $-k_{x}$.
%However, this is an accidental event, and the system
%does not possess mirror symmetry if $g_{0}\neq 0$.

\begin{figure}
\includegraphics[width=2.6in]{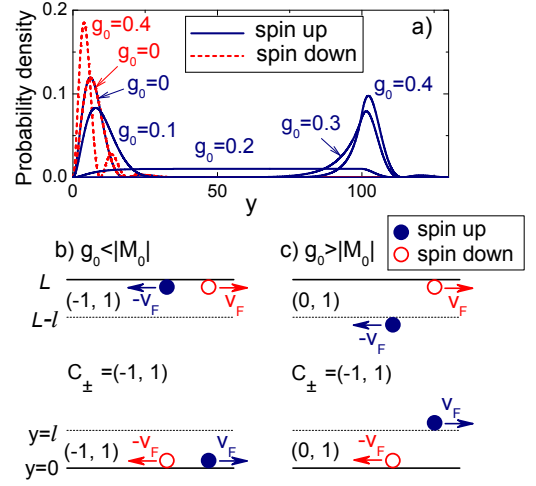}
\caption{(a) $\vert\phi_{+}(k_x,y)\vert^2$ and
$\vert\phi_{-}(k_x,y)\vert^2$ at $k_x=0$ as functions of $y$ for
different $g_{0}$, where $v_{\mbox{\tiny F}}=1$, $B=-5$, $M_{0}=-0.2$,
and $l=100$ are taken. 
%Here, $\vert\phi_{-}(k_x,y)\vert^2$ is plotted
%only for two values of $g_{0}$, as its change with $g_{0}$ is
%relatively small. 
(b,c) Spatial distributions of spin Chern numbers
$C_{\pm}$ and edge states for $g_{0}<\vert M_{0}\vert$ and
$g_{0}>\vert M_{0}\vert$.} \label{Fig2}
  \end{figure}
In  Eqs.\ (\ref{phiP}) and (\ref{phiM}), the
$\hat{\mbox{\boldmath{$\tau$}}}$ and
$\hat{\mbox{\boldmath{$\sigma$}}}$ parts of wavefunctions do not
change with varying $g_{0}$. In Fig.\ 2(a), the modulus squared of
spatial wavefunctions $\phi_{+}(0,y)$ and $\phi_{-}(0,y)$ are
plotted as functions of $y$ for several values of $g_{0}$.
Here, the momentum is taken to be dimensionless by properly
choosing the units for $v_{\mbox{\tiny F}}$ and $B$,
and $v_{\mbox{\tiny F}}$ is used as the unit of energy. At
$g_{0}=0$, we have $\vert\phi_{+}(0,y)\vert^2
=\vert\phi_{-}(0,y)\vert^2$, and both lines coincide with each
other, which is required by the TR symmetry at $g_{0}=0$, as
mentioned above. With increasing $g_{0}$, the peak of
$\vert\phi_{-}(0,y)\vert^2$ becomes sharper and closer to
$y=0$. On the contrary, the shape of
$\vert\phi_{+}(0,y)\vert^2$ widens with increasing $g_{0}$, and
spreads across region I at $g_{0}=\vert M_{0}\vert=0.2$. With further
increasing $g_{0}$, $\vert\phi_{+}(0,y)\vert^2$ becomes
localized near $y=l=100$, i.e., the inner boundary of region I.
%This
%transition is directly caused by the sign change of $\xi_{2}(g_{0})$ in
%$\phi_{+}(0,y)$ at $g_{0}=\vert M_{0}\vert$, from positive to negative.  The
%other characteristic lengths in $\phi_{+}(0,y)$ and $\phi_{-}(0,y)$
%remain positive during this process.

The evolution of the edge states with varying $g_{0}$  is further
illustrated in Figs.\ 2(b) and 2(c), and can be understood in terms of
calculated spin Chern numbers. For $g_{0}<\vert M_{0}\vert$, the spin Chern
numbers $C_{\pm}$ in the three regions take the same value $(-1,
1)$. This indicates that the three regions are topologically
equivalent, and can be regarded as a QSH system as a whole. As a
result, the edge states for both up spin ($\tau_z=1$) and down spin
($\tau_z=-1$) appear near the sample boundaries $y=0$ and $y=L$, as shown
in Fig.\ 2(b). For $g_{0}>\vert M_{0}\vert$, the situation is quite different,
because  $C_{\pm}$ undergo a transition at $g_{0}=\vert M_{0}\vert$ from $(-1,
1)$ to $(0, 1)$ in regions I and II, corresponding to a QAH phase.~\cite{spinch4,HgTe2}
 For the spin-down electrons, the
three regions have the same Chern number $C_{-}=1$, and as a whole
are equivalent to a QH system. Therefore, the spin-down edge states
remain localized near $y=0$ and $y=L$. For the spin-up electrons,
region III with $C_{+}=-1$ is a QH system, sandwiched between two
insulators in regions I and II, where $C_{+}=0$. The spin-up edge
states thus shift to their interfaces, namely, $y=l$ and $y=L-l$, as
shown in Fig.\ 2(c).
Comparing Fig.\ 2(c) for $g_0>\vert M_0\vert$ with Fig.\ 2(b) for $g_0<\vert M_0\vert$,
one finds that both systems have very similar edge states,
so as to exhibit the same QSH effect.
%: the lower
%edge contains a right mover with up spin and a left mover with down
%spin, and conversely for the upper edge, so as to exhibit the same
%QSH effect. 
An important difference is that for $g_0>\vert M_0\vert$, the
counterflows of electrons at the lower (upper) edge are spatially
separated into two different ``lanes'' located at $y=0$ ($y=L$) and
$y=l$ ($y=L-l$), which provides an essential
protection for the edge states against backscattering
from symmetry-breaking random potential. 
%As a result, the QSH state for
%$g_{0}>\vert M_{0}\vert$ should be more stable than that for
%$g_{0}<\vert M_{0}\vert$.

\begin{figure*}
\includegraphics[width=4.5in]{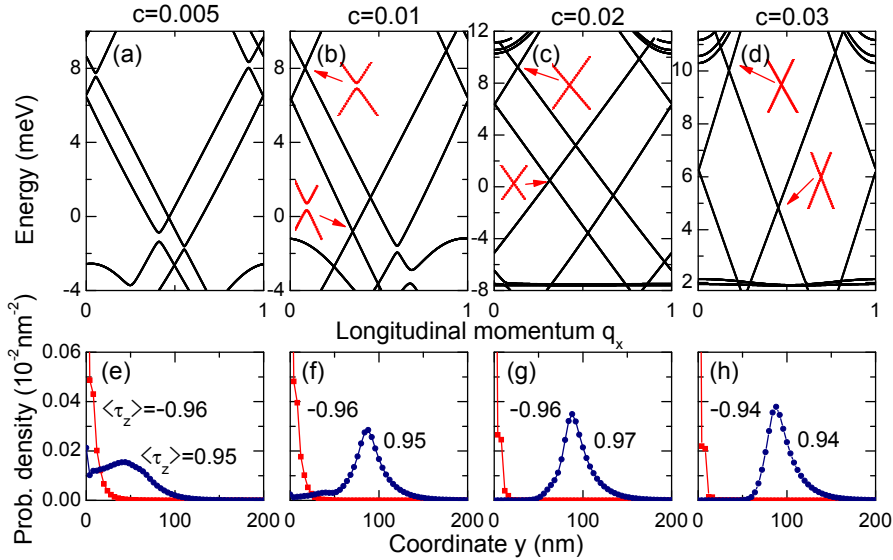}
\caption{(a-d) Calculated eigenenergies for four different
doping concentrations as functions of momentum $q_{x}$ (in units of $2\pi/L_x$).
(e-h) Corresponding
probability density distributions
of the edge states at $E_{\mbox{\tiny F}}=4$meV, which is normalized
in the $L_{x}\times L$ sample. The spin
polarizations $\langle\hat{\tau}_z\rangle$ of the edge states are indicated in (e-h).
} \label{Fig3}
  \end{figure*}
We have shown how the QSH effect can be strengthened by doping of
magnetic atoms near the edges of a QSH system. Now we consider a
more realistic model for HgTe quantum wells doped with Mn atoms
(Hg$_{1-c}$Mn$_c$Te), for which Hamiltonian $H_{0}$ is still given
by Eq.\ (1). However, we take into account the fact that the doped
Mn atoms are spatially randomly distributed, and their spins may not
be fully aligned, so that $H_{1}$ is taken to be
\begin{equation}
H_{1}=-\frac{1}{\pi\lambda^2}\sum_{\alpha=0}^{N_{\mbox{\tiny Mn}}-1}(j_{0}\hat{\sigma}_{z}
+j_{1})\hat{\mbox{\boldmath{$\tau$}}}
\cdot{\bf s}_{\alpha}\exp\left(-\vert {\bf r}-{\bf R}_{\alpha}\vert^2/\lambda^2
\right)\ .
\end{equation}
Here, factor $(j_{0}\hat{\sigma}_{z} +j_{1})$ accounts for different
electron-spin interaction strengths in the electron and hole bands,
${\bf r}$ is the electron coordinate operator, ${\bf R}_{\alpha}$
the position of the $\alpha$-th Mn atom, and $N_{\mbox{\tiny Mn}}$ the total
number of the Mn atoms. The local spins ($S=5/2$) of the Mn atoms
are treated as classical vectors, and ${\bf s}_{\alpha}$ is a unit
vector in the direction of the local spin of the $\alpha$-th Mn
atom.  The distribution of the orientations of the local spins is
assumed to be Boltzmann-Maxwell-like~\cite{PRLSheng} $f({\bf
s}_{\alpha}) \propto e^{-\eta\cos\theta_{\alpha}}$, where
$\theta_{\alpha}$ is the polar angle of ${\bf s}_{\alpha}$, and
$\eta$ is a parameter which can be related to the ratio of
magnetization $M$ to saturated magnetization $M_{s}$:
$M/M_s=-\langle\cos\theta_{\alpha}\rangle=\mbox{coth}(\eta)-1/\eta$.
Therefore, for a given ratio $M/M_{s}$, the distribution is fully
determined. We will set $M=M_{s}/3$, for which the local spins are
randomly oriented to a large degree. As a result, $H_1$ given by
Eq.\ (5) not only provides an exchange field, but also acts as a
scattering potential of magnetic impurities. If one makes the
mean-field approximation, by replacing ${\bf s}_{\alpha}$ with its
average $\langle{\bf s}_{\alpha}\rangle=-(M/M_{s}){\bf e}_{z}$ and
averaging Eq.\ (5) over a random distribution of ${\bf R}_{\alpha}$,
Eq.\ (5) recovers $H_{1}=(g_{0}\hat{\sigma}_{z}
+g_{1})\hat{\tau}_{z}$, where $g_{0}=j_{0}cM/M_{s}a_{0}^2$ and
$g_{1}=j_{1}cM/M_{s}a_{0}^2$ with $a_{0}$ as the lattice constant.
By using the known expressions for mean-field parameters $g_{0}$ and
$g_{1}$,~\cite{HgTe2} we get $j_{0}=464$meV$\cdot$nm$^2$ and
$j_{1}=286$meV$\cdot$nm$^2$, which are independent of $M/M_{s}$ and
doping concentration $c$, as $g_{0}$ and $g_{1}$ are proportional to
$cM/M_{s}$. The coupling range $\lambda$ is set to be
$10$nm.~\cite{couplingRange}

The other parameters of the model
are taken from Ref.~\cite{QWparameter} $v_{\mbox{\tiny F}}=364.5$meV$\cdot$nm, $B=-686$meV$\cdot$nm$^2$,
$D=-512$meV$\cdot$nm$^2$, and $M_0=-10$meV, corresponding to a HgTe quantum well
of thickness $7.0$nm. We consider a sample having the strip geometry shown
in Fig.\ (1), with linear sizes
$L_{x}=80$nm and $L=560$nm in the $x$ and $y$ directions.
%The sample is curled  in the $x$ direction to form a loop,
%inserted with a magnetic flux $\phi$, similar to the setup used in the
%Laughlin gedanken experiment.~\cite{laughlin}
The doped Mn atoms are randomly distributed in
regions of width $l=80$nm near the two edges, with potential
described by Eq.\ (5).
We employ the supercell algorithm,~\cite{supercell} in which
the $L_x\times L$ sample (supercell) is duplicated
along the $x$ direction to form a superlattice.
A tight-binding model on square
meshes is constructed, which recovers the
form of Eq.\ (1) in the continuum limit.
The eigenenergies of the superlattice
as functions of the longitudinal
momentum~\cite{note2} $q_{x}$ are calculated
by exact diagonalization, and the result for four
different doping concentrations is plotted in Figs.\ 3(a-d).
The mesh size is set to be $4$nm, and good convergence is
verified with smaller mesh sizes.
For $c=0.005$, apparent
energy gaps exist in the edge state spectrum,
indicating the occurrence of backscattering.
With increasing $c$ to $0.01$, the energy gaps decrease but
remain finite. With further increasing $c$ to $0.02$ and $0.03$,
the energy gaps essentially vanish, an indication of
quenching of backscattering.
%In this case, electron transport
%through the edge channels is nondissipative, and the
%presence of weak disorder has no influence on it.

%The evolution of eigenenergies with $\phi$
%shown in Figs.\ 3(a-d) can also be considered as the
%energy spectrum of an infinitely long strip made by periodic duplications
%of the above $L_x\times L$ sample along the $x$ direction,
%for which $\phi$ is equivalent to
%the longitudinal momentum (times $L_{x}$).
At a given Fermi energy
$E_{F}=4$meV, the typical spatial probability distributions of the
edge states on an arbitrarily chosen cross-section of the sample for
different doping concentrations are plotted in Figs.\ 3(e-h), in
which only the profile on one side of the sample is shown. For
$c=0.005$, the spin-up and spin-down polarized edge states
both located at $y=0$ have a large spatial overlap. For $c=0.01$, while
the spin-down edge state remains to be near $y=0$, the spin-up edge
state moves away from $y=0$ toward $y=l=80$nm. However, an
appreciable overlap still exists between the two edge states, which
is the origin of finite energy gaps for the edge state spectra in
Figs.\ 3(a) and 3(b). With increasing $c$ to $0.02$ and $0.03$,
the spin-up edge state is peaked at $y=l$, and
there is no longer overlap between the spin-up and spin-down edge
states, as shown in Figs.\ 3(g) and 3(h). This accounts for the
vanishing energy gaps of the edge states shown in Figs.\ 3(c) and
3(d). %These numerical results are in good agreement with the
%physical picture established in the discussion above, and further
%support the feasibility of the present proposal for robust QSH
%effect. 
In conclusion, we  have shown that robust QSH effect can be
realized by placing random Mn impurities on the edge strips. In
principle this leads to conflicting effects: on one hand the induced
exchange field pulls apart the opposite spin edge channels, but on the
other hand it provides a mechanism for backscattering by magnetic
impurities. The present calculation, for realistic parameters of HgTe
quantum wells, implies that the combined effect is stabilizing the QSH
effect rather than vice versa.

%To
%show this point, we introduce a specific perturbation violating the
%TR symmetry. Since the two edge states given by Eqs.\ (3) and (5)
%have opposite $\tau_z$, we include an in-plane Zeeman field
%throughout the sample with potential
%\begin{equation}
%V=\gamma\hat{\tau}_y\ ,
%\end{equation}
%which represents
%a ``maximal'' coupling between the two edge states
%near the degenerate point.  A tight-binding model on a square
%lattice is constructed for the strip geometry, which recovers the
%form of Eq.\ (1) in the continuum limit. The energy spectrum is
%calculated numerically from the tight-binding model.

%\section{ACKNOWLEDGMENTS}

This work is supported by the State Key Program for Basic Researches
of China under Grants Nos. 2009CB929504 (LS),
2011CB922103, and 2010CB923400 (DYX), the National Natural Science
Foundation of China under Grant Nos. 11225420, 11074110 (LS),
11174125, 11074109, 91021003 (DYX),
and a project funded by the PAPD
of Jiangsu Higher Education Institutions. We also thank the US NSF Grants
No. DMR-0906816 and No. DMR-1205734, and
Princeton MRSEC Grant No. DMR-0819860 (DNS).

%ÎÄÏײ¿·Ö

%\end{CJK}
\end{document}